\documentclass[aps,prd,preprint,groupedaddress]{revtex4}

\usepackage{epsfig}

\begin{document}

\def\lp{\left. }
\def\rp{\right. }
\def\lr{\left( }
\def\rr{\right) }
\def\le{\left[ }
\def\re{\right] }
\def\lg{\left\{ }
\def\rg{\right\} }
\def\lb{\left| }
\def\rb{\right| }

\def\go{\tilde{g}}
\def\mg{m_{\go}}

\def\msQ {m_{\tilde{Q}}}
\def\msD {m_{\tilde{D}}}
\def\msU {m_{\tilde{U}}}
\def\msS {m_{\tilde{S}}}
\def\msC {m_{\tilde{C}}}
\def\msB {m_{\tilde{B}}}
\def\msT {m_{\tilde{T}}}
\def\msusy{m_{\rm SUSY}}

\def\sq  {\tilde{q}}
\def\sql {\tilde{q}_L}
\def\sqr {\tilde{q}_R}
\def\ms  {m_{\sq}}
\def\msql{m_{\tilde{q}_L}}
\def\msqr{m_{\tilde{q}_R}}

\def\st  {\tilde{t}}
\def\stl {\tilde{t}_L}
\def\str {\tilde{t}_R}
\def\mstl{m_{\stl}}
\def\mstr{m_{\str}}
\def\sta {\tilde{t}_1}
\def\stb {\tilde{t}_2}
\def\msta{m_{\sta}}
\def\mstb{m_{\stb}}
\def\thst{\theta_{\tilde{t}}}

\def\sb  {\tilde{b}}
\def\sbl {\tilde{b}_L}
\def\sbr {\tilde{b}_R}
\def\msbl{m_{\sbl}}
\def\msbr{m_{\sbr}}
\def\sba {\tilde{b}_1}
\def\sbb {\tilde{b}_2}
\def\msba{m_{\sba}}
\def\msbb{m_{\sbb}}
\def\thsb{\theta_{\tilde{b}}}

\newcommand{\sfa}{\tilde{f}}

\newcommand{\SLASH}[2]{\makebox[#2ex][l]{$#1$}/}
\newcommand{\kslash}{\SLASH{k}{.15}}
\newcommand{\pslash}{\SLASH{p}{.2}}
\newcommand{\qslash}{\SLASH{q}{.08}}

\def\d  {{\rm d}}
\def\eps{\varepsilon}

\def\beq{\begin{equation}}
\def\eeq{\end{equation}}
\def\bea{\begin{eqnarray}}
\def\eea{\end{eqnarray}}

\preprint{DESY 02-119}
\preprint{hep-ph/0208212}
\title{Gluino Pair Production at Linear \boldmath $e^+e^-$ Colliders}
\author{Stefan Berge}
\affiliation{{II.} Institut f\"ur Theoretische Physik, Universit\"at Hamburg,
             Luruper Chaussee 149, D-22761 Hamburg, Germany}
\author{Michael Klasen}
\email[]{michael.klasen@desy.de}
\affiliation{{II.} Institut f\"ur Theoretische Physik, Universit\"at Hamburg,
             Luruper Chaussee 149, D-22761 Hamburg, Germany}
\date{\today}
\begin{abstract}
We study the potential of high-energy linear $e^+e^-$ colliders for the
production of gluino pairs within the Minimal Supersymmetric Standard Model
(MSSM). In this model, the process $e^+e^-\to\go\go$ is mediated by
quark/squark loops, dominantly of the third generation, where the mixing of
left- and right-handed states can become large. Taking into account realistic
beam polarization effects, photon and $Z^0$-boson exchange, and current mass
exclusion limits, we scan the MSSM parameter space for various $e^+e^-$
center-of-mass energies to determine the regions, where gluino production
should be visible.
\end{abstract}
\maketitle


\section{Introduction}
\label{sec:1}

Supersymmetry (SUSY) is generally considered to be one of the most promising
extensions of the Standard Model (SM) of particle physics. Its attractive
features include the cancellation of quadratic divergences in the Higgs sector,
which implies that the soft SUSY breaking masses of the (yet unobserved)
superpartners of the SM particles can not be much greater than the electroweak
scale. If SUSY is indeed responsible for the stabilization of this
scale against the Planck scale, supersymmetric particles should therefore be
discovered either at Run II of the Fermilab Tevatron
\cite{Carena:2000yx,Abel:2000vs,
Culbertson:2000am,Ambrosanio:1998zf,Allanach:1999bf} or at the CERN LHC
\cite{:1999fr,Abdullin:1998pm}. In particular, the strongly coupling squarks
and gluinos should be copiously produced at hadron colliders and lead to first
measurements of their masses and production cross sections
\cite{Beenakker:1996ch}. Precision measurements of masses, mixings, quantum
numbers, and couplings must, however, be performed in the clean environment of
a future linear $e^+e^-$ collider because of the large hadronic SM background
and theoretical scale and parton density uncertainties at the Fermilab
Tevatron and CERN LHC.
For example, in $e^+e^-$ annihilation the center-of-mass energy of the
collision is exactly known, and threshold energy scans allow for a precise mass
determination of pair-produced SUSY particles. It will then be possible to
establish whether the masses and couplings of the electroweak gauginos and of
the gluino are indeed related, as expected. A global analysis should
ultimately lead to a reconstruction of the SUSY breaking model and its
parameters. Along these lines, detailed studies have recently been performed
for squarks, sleptons, charginos, and neutralinos
\cite{Aguilar-Saavedra:2001rg}, but not for gluinos, the reason being that
gluino pairs are produced at $e^+e^-$ colliders only at the one-loop level,
while all other sparticles are produced at tree level. At tree level, gluinos
can be produced in pairs only in association with two quarks
\cite{Campbell:1981hf}, or they are produced singly in association with a
quark and a squark \cite{Schiller:kq,Brandenburg:2002ff}. Both processes
result in
multi-jet final states, where phase space is limited and gluinos may be hard
to isolate.

In the Minimal Supersymmetric Standard Model (MSSM) \cite{Nilles:1983ge,
Haber:1984rc}, the exclusive production of gluino pairs in $e^+e^-$
annihilation is mediated by $s$-channel photons and $Z^0$-bosons,
which couple to the gluinos via triangular quark and squark loops. In earlier
studies of this process only the very low center-of-mass energy region
($\sqrt{s}=20$ GeV) with pure photon exchange and no squark mixing
\cite{Nelson:1982cu} or $Z^0$-boson decays into light ($\mg < m_Z/2$)
\cite{Kane:xp,Campbell:kf} and very light ($\mg=3\,...\,5$ GeV)
\cite{Kileng:1994kw,Djouadi:1994js} gluinos have been considered. Some authors
have presented results only for $\mg=0$ GeV and outdated top quark
masses of $20\,...\,50$ GeV \cite{Kane:xp,Djouadi:1994js}, while others have
neglected the mixing of left- and right-handed squark interaction eigenstates
into light and heavy mass eigenstates \cite{Nelson:1982cu,Kane:xp,Campbell:kf},
which turns out to control the production cross section to a large extent.

It is the aim of this Article to study the potential of high-energy
linear $e^+e^-$ colliders for the production of gluino pairs within the MSSM.
Taking into account realistic beam polarization effects, photon and $Z^0$-boson
exchange, and current mass exclusion limits, we scan the MSSM parameter space
for various $e^+e^-$ center-of-mass energies to determine the regions, where
gluino production should be visible. Furthermore we clarify the theoretical
questions of the relative sign between the two contributing triangular Feynman
diagrams, of the possible presence of an axial vector anomaly, and the
conditions for vanishing cross sections -- three related issues, which have so
far been under debate in the literature.
The remainder of this Paper is organized as follows: In Sec.\ \ref{sec:2},
we present our analytical results and compare them with existing results in
the literature. Various numerical cross-sections for gluino pair production at
future high-energy linear $e^+e^-$ colliders are computed and discussed in
Sec.\ \ref{sec:3}, and Sec.\ \ref{sec:4} contains our Conclusions. Our
conventions for squark mixing are defined in App.\ \ref{sec:a}, and a
summary of all relevant Feynman rules is given in App.~\ref{sec:b}.

\section{Analytical results}
\label{sec:2}

\begin{figure}
 \centering
 \epsfig{file=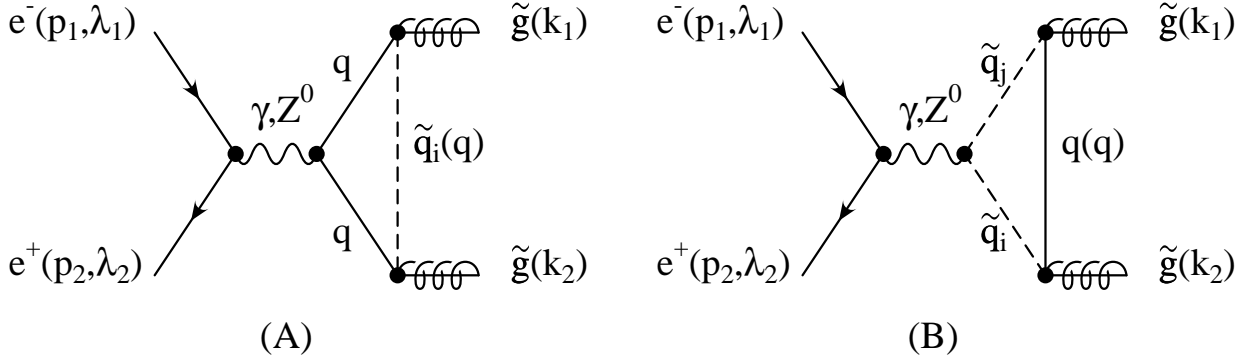,bbllx=38pt,bblly=349pt,bburx=314pt,bbury=431pt,%
  width=\columnwidth}
 \caption{\label{fig:1}Feynman diagrams for gluino pair production in
 electron-positron annihilation. The exchanged photons and $Z^0$-bosons
 couple to the produced gluinos through triangular $qq\sq_i$ (A) and $\sq_i
\sq_j q$ (B) loops with flavor flow in both directions.}
\end{figure}

The scattering process
\beq
 e^-(p_1,\lambda_1)e^+(p_2,\lambda_2)\to\go(k_1)\go(k_2)
\eeq
with incoming electron/positron momenta $p_{1,2}$ and helicities
$\lambda_{1,2}$ and outgoing gluino momenta $k_{1,2}$ proceeds through
the two Feynman diagrams A and B in Fig.\ \ref{fig:1}
%
%
with $s$-channel photon and $Z^0$-boson exchange and triangular quark and
squark loops. Higgs boson exchange is not considered due to the negligibly
small electron Yukawa coupling, but it could well be relevant at
muon colliders. The process occurs only at the one-loop level, since the gluino
as the superpartner of the gauge boson of the strong interaction couples
neither directly to leptons nor to electroweak gauge bosons. Taking into
account chiral squark mixing (see App.\ \ref{sec:a}) and using the Feynman
rules in App.\ \ref{sec:b}, we decompose the corresponding scattering amplitude
\beq
 {\cal M} = \sum_{V=\gamma,Z^0} L_\mu^V i D_V^{\mu\nu} G_\nu^V
\eeq
into the lepton current
\beq
 L_\mu^V = \bar{v}(p_2,\lambda_2) \le-ie\gamma_\mu (v_e^V-a_e^V\gamma_5)\re
 u(p_1,\lambda_1),
\eeq
the photon and $Z^0$-boson propagators
\beq
 iD^{\mu\nu}_V = {-i g^{\mu\nu} \over s-m_V^2+i\eta}\ , \quad V = \gamma,\,
 Z^0,
\eeq
which depend on the squared center-of-mass energy $s=(p_1+p_2)^2$, and the
gluino current
\bea
 G_\nu^V &=& -e \bar{u}(k_2) \sum_q \le
  \sum_i     \lr i\Gamma   ^a_{i,1} A ^{i ,V}_\nu i\Gamma   ^b_{i,2}
              +  i\Gamma'\,^a_{i,2} \tilde{A}^{i ,V}_\nu i\Gamma'\,^b_{i,1}
 \rr \rp  \label{eq:gten1} \\ && \lp \hspace*{19mm}
 +\sum_{i,j} \lr i\Gamma   ^a_{i,1} \Gamma^{ij,V}_q B   ^{ij,V}_\nu i\Gamma ^b_{j,2}
              +  i\Gamma'\,^a_{i,2} \Gamma^{ji,V}_q \tilde{B}^{ij,V}_\nu i\Gamma'^b_{j,1}
 \rr \re v(k_1) \nonumber
\eea
with squark mass eigenstates $i,j=\{1,2\}$,
\beq
 A^{i ,V}_\nu = \int {\d^Dq \over (2\pi)^D} \mu^{4-D}
 {(\qslash+\kslash_2+m_q)\gamma_\nu (v_q^V-a_q^V\gamma_5)
  (\qslash-\kslash_1+m_q) \over
  (q^2-m_i^2+i\eta)[(q-k_1)^2-m_q^2+i\eta][(q+k_2)^2-m_q^2+i\eta]}
 \label{eq:loopa}
\eeq
and
\beq
 B^{ij,V}_\nu = \int {\d^Dq \over (2\pi)^D} \mu^{4-D}
 {(\qslash-m_q)(2q - k_1 +k_2)_\nu \over
  (q^2-m_q^2+i\eta)[(q-k_1)^2-m_j^2+i\eta][(q+k_2)^2-m_i^2+i\eta]}
 \label{eq:loopb}
\eeq
and the quark flavor $q$ flowing both ways in the corresponding diagrams A and
B, so that $\tilde{A}^{i,V}_\nu=A^{i,V}_\nu(v_q^V\to-v_q^V),\,\tilde{B}^{ij,V}_\nu=
-B^{ij,V}_\nu,$ and $\Gamma'=C\Gamma^{\rm T}C^{-1}=\Gamma$ \cite{Denner:kt}.

Eq.\ (\ref{eq:gten1}) can be simplified
with the Dirac equation and the
anti-commutation relations for Dirac matrices, and the tensor loop integrals in
Eqs.\ (\ref{eq:loopa}) and (\ref{eq:loopb}) can be expressed through the
standard coefficient functions $C_{k(l)}$ of the metric tensor $g_{\mu\nu}$
and tensors constructed from the outgoing gluino momenta $k_{1,\mu}$ and
$k_{2,\mu}$ \cite{Passarino:1978jh,Denner:kt}. The gluino current then reduces
to
\beq
 G_\nu^V = i e {\alpha_s \over 2 \pi} {\delta^{ab} \over 2}
           \bar{u}(k_2) \gamma_\nu \gamma_5 v(k_1)
           \sum_q (A^V_q + B^V_q)
\eeq
with
\bea
 A^V_q &=& \sum_i \le
   C^{qi}_0 (m^2_q           a^-_{qiV}
            -m^2_{\tilde{g}} a^+_{qiV}
            +2 m_q m_{\tilde{g}} \hat{a}_{qiV})
 + C^{qi}_1    4 m_{\tilde{g}} (m_q \hat{a}_{qiV} - m_{\tilde{g}} a^+_{qiV})
 \rp \\ & & \hspace{4.2mm} \lp
 + C^{qi}_{00} (2-D)\, a^+_{qiV}
 - C^{qi}_{11} 2 m^2_{\tilde{g}} a^+_{qiV}
 + C^{qi}_{12} \lr s-2 m^2_{\tilde{g}}\rr a^+_{qiV}\re, \nonumber \\
 B^V_q &=& \sum_{i,j} C^{qij}_{00} 2 b_{qijV},
\eea
where $C^{qi}_{k(l)}=C_{k(l)}(m^2_{\tilde{g}},s,m^2_{\tilde{g}},
m^2_{\tilde{q}_i},m^2_q,m^2_q)$ and $C^{qij}_{00}=C_{00}(m^2_{\tilde{g}},s,
m^2_{\tilde{g}},m^2_q,m^2_{\tilde{q}_j},m^2_{\tilde{q}_i})$ are massive
(infrared-finite) three-point functions and $C_1^{qi}=C_2^{qi}$ and
$C_{11}^{qi}=C_{22}^{qi}$ in diagram A.
\bea
 a^{\pm}_{qiV} &=& v^V_q (S^{q}_{i1}S^{q\ast}_{i1}-S^{q}_{i2}S^{q\ast}_{i2})
                   \pm a_q^V, \nonumber \\
 \hat{a}_{qiV} &=& a^V_q (S^{q}_{i1}S^{q\ast}_{i2}+S^{q}_{i2}S^{q\ast}_{i1})
                            ,\ {\rm and}  \label{eq:abcoup} \\
 b_{qijV}      &=&        S^{q    }_{i1}S^{q\ast}_{j1}\Gamma_q^{ij,V}
                -         S^{q\ast}_{i2}S^{q    }_{j2}\Gamma_q^{ji,V} \nonumber
\eea
are combinations of vector ($v^V_q$), axial vector ($a^V_q$), and
derivative couplings ($\Gamma_q^{ij,V}$), and elements of the squark mixing
matrix
$S$. Pairs of identical (Majorana) gluinos are therefore produced by a parity
violating axial vector coupling induced by mass differences between the chiral
squarks and the axial vector coupling $a_q^Z$ of the $Z^0$-boson. The
(mass-independent) ultraviolet singularities contained in the
$C_{00}$-functions cancel among $A_q^V$ and $B_q^V$ in \mbox{$D=4-2\eps$}
dimensions.
As we have checked explicitly (even for complex squark mixing matrices),
adding the two amplitudes induces not only a
cancellation of the ultraviolet singularities and of the logarithmic
dependence on the scale parameter $\mu$ introduced in Eqs.\ (\ref{eq:loopa})
and (\ref{eq:loopb}) in order to preserve the mass dimension of the loop
integrals, but also a destructive interference of the finite remainders.
This happens separately for each weak isospin partner, as is to be expected
for triangular loop diagrams involving one axial vector and two scalar (not
vector) couplings and no closed fermion loop.

The (finite) total cross section for incoming electrons/positrons with
helicities $\lambda_{1,2}=\pm1/2$ is then
\beq
 \sigma_{\lambda_1\lambda_2} (s) = {\alpha_e^2\alpha_s^2 (N_C^2-1) \beta^3 s
 \over 24 \pi}\sum_{V_1,V_2} \le {Q^{V_1 V_2}_{\lambda_1\lambda_2} \over
 (s-m_{V_1}^2)(s-m_{V_2}^2)} \sum_q (A^{V_1}_q+B^{V_1}_q)
 (A^{V_2}_q+B^{V_2}_q)^\ast \re
 \label{eq:polxsec}
\eeq
with
\beq
 Q^{V_1V_2}_{\lambda_1\lambda_2} = (v_e^{V_1}v_e^{V_2}+a_e^{V_1}a_e^{V2})
 (1-4\lambda_1\lambda_2)-(v_e^{V_1}a_e^{V2}+v_e^{V_2}a_e^{V_1})(2\lambda_1
 -2\lambda_2),
\eeq
color factor $N_C=3$, and gluino velocity $\beta=\sqrt{1-4m^2_{\tilde{g}}/s}$,
which contains the expected factors of $\beta^3$ and $s$ for $P$-wave
production of two spin-1/2 Majorana fermions. The distribution in the
center-of-mass scattering angle $\theta$,
\beq
 {\d\sigma_{\lambda_1\lambda_2}\over\d\Omega}(s)={3\over 8\pi}
 (1+\cos^2\theta)\sigma_{\lambda_1\lambda_2}(s),
\eeq
is independent of the gluino mass and has to be integrated over just one
hemisphere, since the two final state particles are identical
\cite{Kileng:1994kw,Petcov:1984nf}. As a consequence, the forward-backward
asymmetry vanishes for Majorana fermions, but not for Dirac fermions.

Our result for diagram A agrees with the unpolarized result
\beq
 \sigma(s)={1\over 4}\sum_{\lambda_{1,2}=\pm 1/2}
 \sigma_{\lambda_1\lambda_2}(s)
\eeq
in Eq.\ (4.5) of Kileng and Osland, if we identify \cite{Kileng:1994kw,
Kileng:2002}
\bea
 C_0   ^{qi } &=& - F^{00}_{qqi},\nonumber\\
 C_1   ^{qi } &=& + F^{01}_{qqi},\nonumber\\
 C_{00}^{qi } &=& - G_{qqi}/2   ,\\
 C_{11}^{qi } &=& - F^{02}_{qqi},\nonumber\\
 C_{12}^{qi } &=& - F^{11}_{qqi},\nonumber
\eea
and reverse the sign of $\hat{b}_q$ to account for opposite conventions of
squark mass eigenstates. However, our result for diagram B disagrees in sign
with Eq.\ (4.5) of Kileng and Osland, if we identify \cite{Kileng:1994kw,
Kileng:2002}
\beq
 C_{00}^{qij} = - G_{ijq}/2.
\eeq
If the sign of diagram B is reversed, the ultraviolet singularities cancel only
after adding the contributions from the two weak isospin partners with
opposite values of $T_q^3$.
We trace this sign discrepancy to the Feynman rules employed in Ref.\
\cite{Kileng:1994kw}, which exhibit a relative minus sign to those in Ref.\
\cite{Haber:1984rc} for the $Z^0$-boson coupling to quarks, but not to
squarks, whereas our Feynman rules (see App.\ \ref{sec:b}) agree with those
in Ref.\ \cite{Haber:1984rc} in the limit of no squark mixing.
Except for the relative sign of diagrams A and B and in the limit of
vanishing gluino mass, we also find agreement with Djouadi and Drees
\cite{Djouadi:1994js}. We confirm, however, the relative sign for diagrams A
and B of Campbell, Scott, and Sundaresan \cite{Campbell:kf}, who found (for
non-mixing chiral squarks of different mass) that the ultraviolet singularities
cancel separately for each weak isospin partner, and that there is no anomaly.
This had also been claimed previously by Kane and Rolnick \cite{Kane:xp} for
chiral squarks of equal mass. In their limit, the cross section depends only on
the weak isospin and not on the charge of the (s)quarks \cite{Kane:xp}, and the
contribution of the photon vanishes \cite{Nelson:1982cu}. For the contribution
of the $Z^0$-boson to vanish, we must have \cite{Kileng:1994kw}
\begin{enumerate}
 \item mass degeneracy in each quark isospin doublet, $m_d=m_u$ etc.,
 \item mass degeneracy in each squark isospin doublet, $m_{\tilde{d}_1}=
  m_{\tilde{d}_2}=m_{\tilde{u}_1}=m_{\tilde{u}_2}$ etc.,
\end{enumerate}
which contradicts the condition $m_q=\ms$ found by Kane and Rolnick
\cite{Kane:xp}. Condition (1) is violated most strongly for the third
generation, as is condition (2) for most SUSY breaking models.

\section{Numerical Results}
\label{sec:3} 

The analytical results presented in the previous Section have been obtained
in two independent analytical calculations. They have been implemented
in compact Fortran computer codes, which depend on the LoopTools/FF library
\cite{Hahn:1998yk,vanOldenborgh:1990yc} for the evaluation of the massive
tensor three-point functions. As a third independent cross-check, we have
recalculated the production of gluino pairs in $e^+e^-$ annihilation with the
computer algebra program FeynArts/FormCalc \cite{Hahn:2001rv} and found
numerical agreement up to 15 digits.

Our calculations involve various masses and couplings of SM particles,
for which we use the most up-to-date values from the 2002 Review of
the Particle Data Group \cite{Hagiwara:pw}. In particular, we evaluate the
electromagnetic fine structure constant $\alpha(m_Z)=1/127.934$ at the mass of
the $Z^0$-boson, $m_Z = 91.1876$ GeV, and calculate the weak mixing angle
$\theta_W$ from the tree-level expression $\sin^2\theta_W=1-m_W^2/m_Z^2$ with
$m_W = 80.423 $ GeV. Among the fermion masses, only the one of the top quark,
$m_t = 174.3$ GeV, plays a significant role due to its large splitting from the
bottom quark mass, $m_b = 4.7$ GeV, while the latter and the charm quark mass,
$m_c = 1.5$ GeV, could have been neglected like those of the three light quarks
and of the electron/positron. The strong coupling constant is evaluated at the
gluino mass scale from the one-loop expression with five active flavors and
$\Lambda_{\rm LO}^{n_f=5} = 83.76$ MeV, corresponding to $\alpha_s(m_Z) =
0.1172$. A variation of the renormalization scale by a factor of four about the
gluino mass results in a cross section uncertainty of about $\pm25$ \%. Like
the heavy top quark, all SUSY particles have been decoupled from the running
of the strong coupling constant.

We work in the framework of the MSSM with conserved $R$- (matter-) parity,
which represents the simplest phenomenologically viable model, but which is
still sufficiently general to not depend on a specific SUSY breaking mechanism.
Models with broken $R$-parity are severely restricted by the non-observation
of proton decay, which would violate both baryon and lepton number
conservation.
We do not consider light gluino mass windows, on which the literature has
focused so far and which may or may not be excluded from searches at fixed
target and collider experiments \cite{Hagiwara:pw}. Instead, we adopt the
current mass limit $\mg\geq 200$ GeV from the CDF \cite{Affolder:2001tc} and
D0 \cite{Abachi:1995ng} searches in the jets with missing energy channel,
relevant for non-mixing squark masses of $\ms\geq 325$ GeV and $\tan\beta = 3$.
Values for the ratio of the Higgs vacuum expectation values, $\tan\beta$, below
2.4 are already excluded by the CERN LEP experiments \cite{lhwg:2001xx}. If not
stated otherwise, we will present unpolarized cross sections for a $\sqrt{s}=
500$ GeV linear $e^+e^-$ collider like DESY TESLA, gluino masses of $\mg=200$
GeV, and squark masses $\ms\simeq\msQ=\msD=\msU=\msS=\msC=\msB=\msT=
\msusy=325$ GeV.
We will consider two cases of large squark mass splittings: I.) On the one
hand, the masses of the superpartners of left- and right-handed
quarks need not be equal to each other. In this scenario we will vary the
right-handed up-type squark mass parameters
$m_{\tilde{U},\tilde{C},\tilde{T}}$ between 200 and 1500 GeV. II.) On the
other hand, the superpartners of the heavy quarks can mix into light and heavy
mass eigenstates (see App.\ \ref{sec:a}). This alternative is restricted by
the CERN LEP limits on the light top and bottom squark masses, $\msta\geq 100$
GeV and $\msba\geq 99$ GeV \cite{lswg:2002xx}, and on SUSY one-loop
contributions \cite{Barbieri:1983wy,Drees:1990dx,Chankowski:1993eu} to the
$\rho$-parameter, $\rho_{\rm SUSY} < 0.0012$ \cite{Hagiwara:pw}.
In this case we assume the maximally
allowed top squark mixing with $\thst=45.2^\circ$, $\msta=110$ GeV, and
$\mstb=506$ GeV, which can be generated by choosing appropriate values for the
Higgs mass parameter, $\mu=-500$ GeV, and the trilinear top squark coupling,
$A_t=534$ GeV. For small values of $\tan\beta$, mixing in the bottom squark
sector remains small, and we take $\theta_{\sb}=0^\circ$.
Although the absolute magnitude of the cross section depends strongly on the
gluino mass and collider energy, the relative importance of the different
contributions is very similar also for higher gluino masses and collider
energies.

\begin{figure}
 \centering
 \epsfig{file=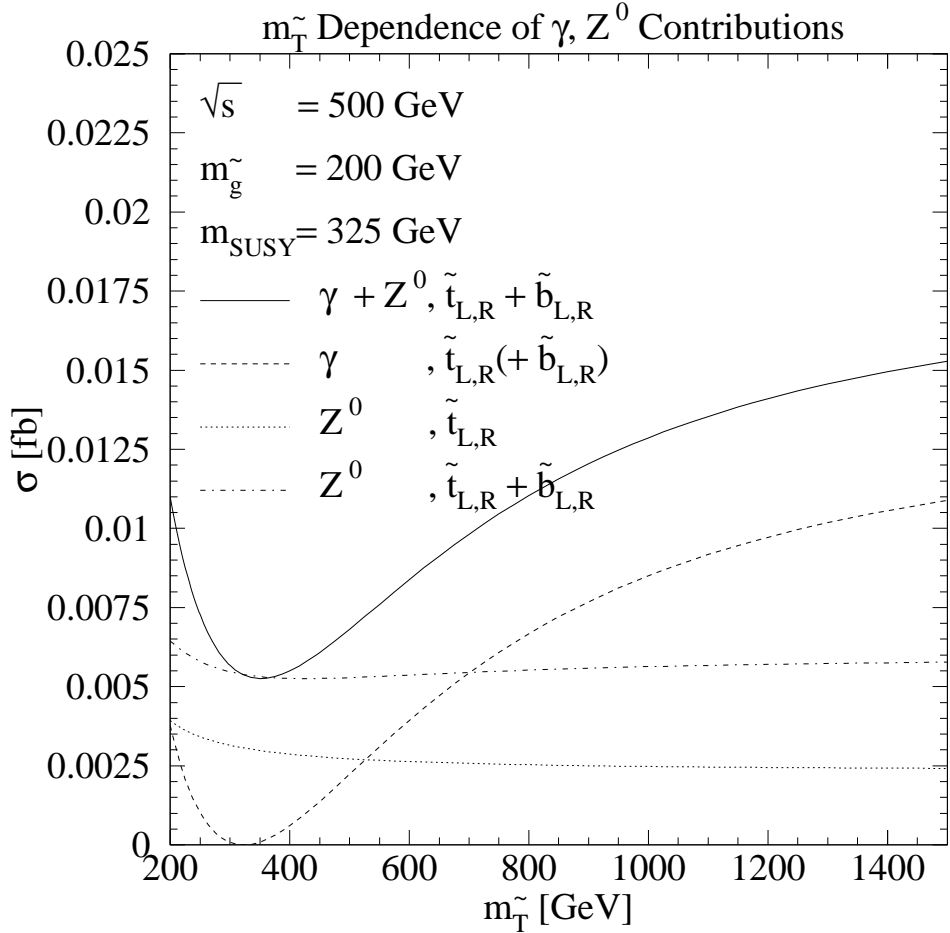,width=0.8\columnwidth}
 \caption{\label{fig:4}Dependence of the photon and $Z^0$-boson contributions
 to the process $e^+e^-\to\go\go$ on the right-handed top squark mass parameter
 $\msT$. The photon contribution (dashed curve) is dominated by top (s)quarks
 and cancels for $\mstl=\mstr$. The $Z^0$-boson contribution from top (dotted
 curve) and bottom squarks (dot-dashed curve) interferes constructively with
 the photon contribution (full curve).}
\end{figure}

First we examine the conditions found in Section \ref{sec:2} for vanishing of
the photon and $Z^0$-boson contributions, restricting ourselves to the
third generation. Since we expect the photon contribution to cancel for equal
left- and right-handed squark masses, we vary the right-handed top squark mass
parameter, $\msT\simeq\mstr$, between 200 and 1500 GeV, but keep $\mstl\simeq
\msQ=\msB=\msusy=325$ GeV fixed (case I), since top and bottom squarks
generally interfere destructively due to their opposite charge and weak isospin
quantum numbers. As can be seen from Fig.\
\ref{fig:4}, the photon contribution cancels indeed for
%
%
$\msT\simeq\mstr=\mstl\simeq\msusy$. This is due to the fact that for photons
$\hat{a}_{qi\gamma}=0$ and $b_{q12\gamma}=b_{q21\gamma}=0$ in Eq.\
(\ref{eq:abcoup}), while unitarity of the squark mixing matrix leads to
$a^\pm_{q1\gamma}=-a^\pm_{q2\gamma}$ and $b_{q11\gamma}=-b_{q22\gamma}$.
Therefore, the photon contributions cancels for all flavors $q$
with equal squark masses. Due
to their charge, top (s)quarks contribute four times as much as bottom
(s)quarks, whose contribution is even more suppressed by the condition $\msbl
\simeq\msbr$. The $Z^0$-boson contribution can never cancel, since $m_t\gg
m_b$, and therefore it depends only weakly on $\msT$, but it can become minimal
for $\msT\simeq\mstr=\mstl=\msbr=\msbl\simeq\msusy$. As $\msT$ gets
significantly larger (or smaller) than $\msusy$, the photon contribution
starts to dominate over the $Z^0$-boson contribution.

\begin{figure}
 \centering
 \epsfig{file=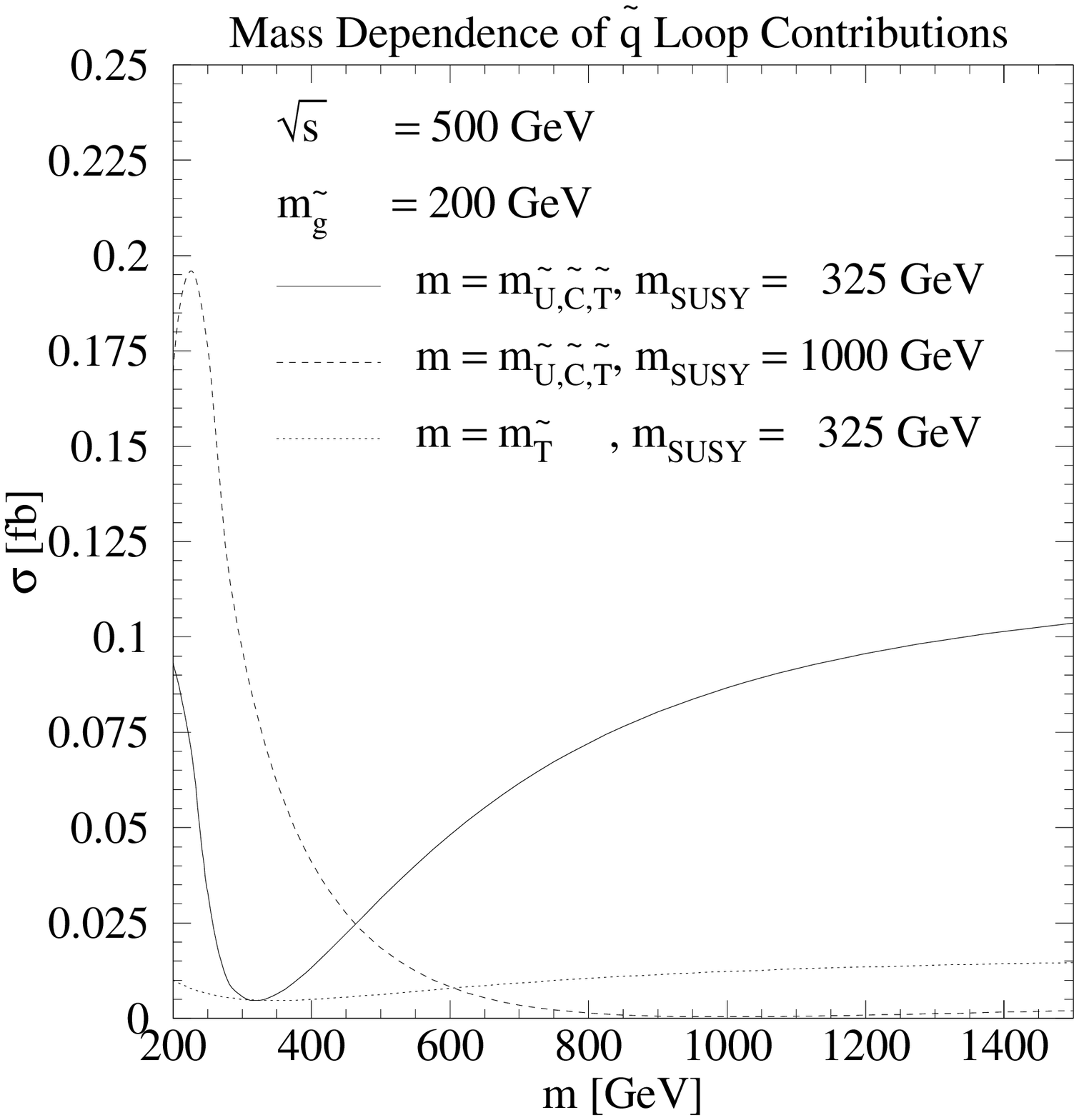,width=0.8\columnwidth}
 \caption{\label{fig:5}Dependence of the $\sq$ loop contributions to the
 process $e^+e^-\to\go\go$ on the right-handed up-type squark mass parameter
 $\msU=\msC=\msT$. When these mass parameters differ simultaneously from
 $\msusy=325$ GeV (full curve) or $\msusy=1000$ GeV (dashed curve), all three
 (s)quark generations contribute significantly to the total cross section, so
 that it becomes much larger than in the case where only $\msT$ is varied
 (dotted curve).}
\end{figure}

If only $\msT$ differs from $\msusy$, the third generation contributes almost
100\% to the total cross section. However, if $\msU=\msC=\msT$ are varied
simultaneously, all three generations contribute to the total cross section,
which can therefore become significantly larger. This is shown in Fig.\
\ref{fig:5}, where (s)quark loop contributions from all three generations
%
%
have been taken into account.

\begin{figure}
 \centering
 \epsfig{file=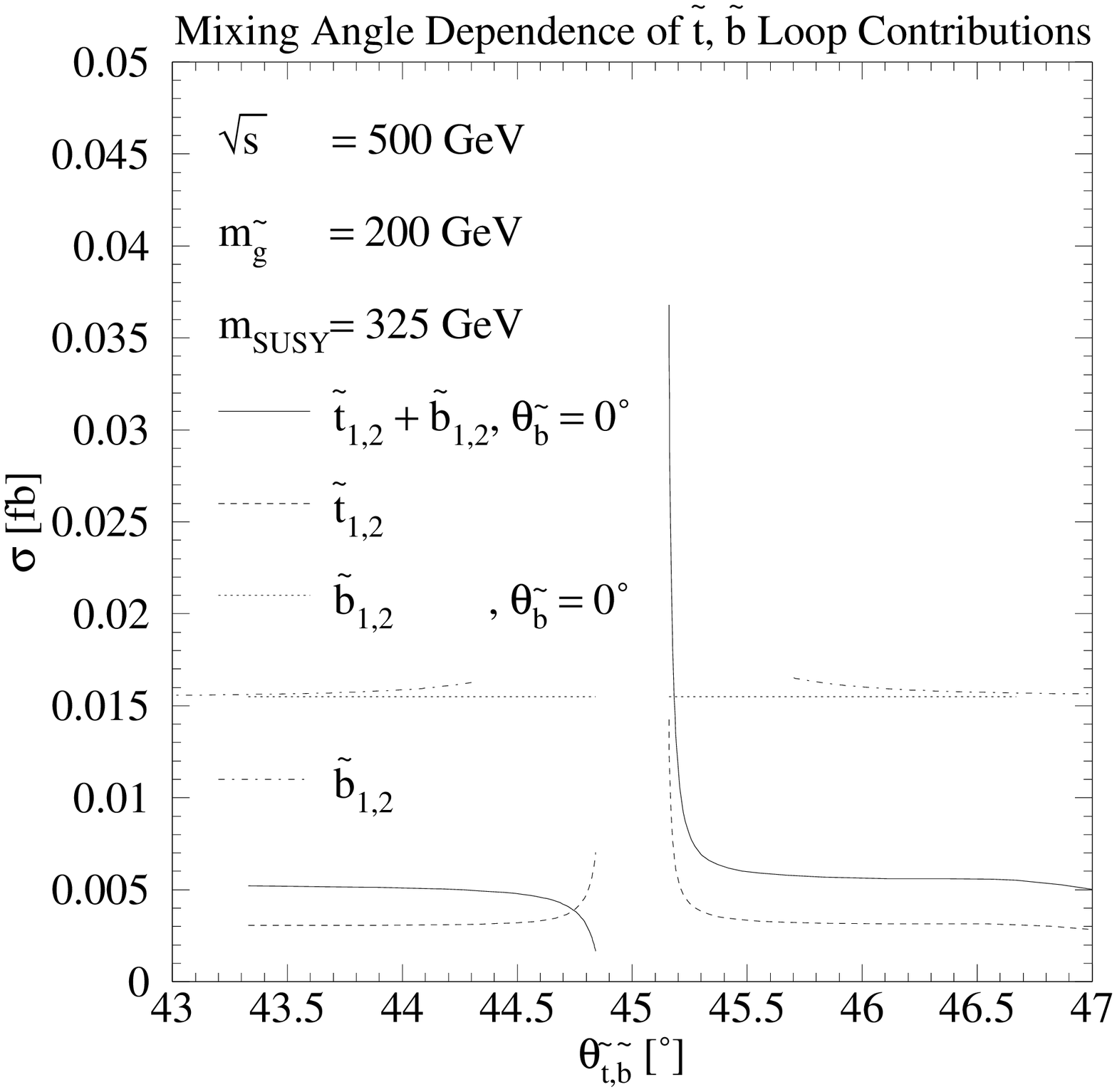,width=0.8\columnwidth}
 \caption{\label{fig:6}Mixing angle dependence of the $\st$ (dashed) and $\sb$
 (dotted) loop contributions to the process $e^+e^-\to\go\go$, which interfere
 destructively (full curve), except for $\thst\simeq 45.2^\circ$, where the
 imaginary parts of the amplitudes interfere constructively. Mixing in the
 $\sb$ sector (dot-dashed curve) enhances the cross section only slightly.}
\end{figure}

When $\msU=\msC=\msT=\msusy$ and large mass splittings are generated only by
mixing in the top squark sector (case II), photon contributions are suppressed
by more than two orders of magnitude. Fig.\ \ref{fig:6} shows that the
%
%
$Z^0$-boson contributions from top and bottom squarks interfere destructively
due to opposite values of their weak isospin quantum numbers, except for
$\theta_{\st}\simeq 45.2^\circ$, where the imaginary parts of the amplitudes
interfere constructively. It is therefore advantageous to keep the bottom
squark mass splitting small. As is also evident from Fig.\ \ref{fig:6}, mixing
in the bottom squark sector is of little importance. Note that
the central region with maximal top/bottom squark mixing is excluded by the
CERN LEP limits on $\msta$, $\msba$, and the $\rho$-parameter.

\begin{figure}
 \centering
 \epsfig{file=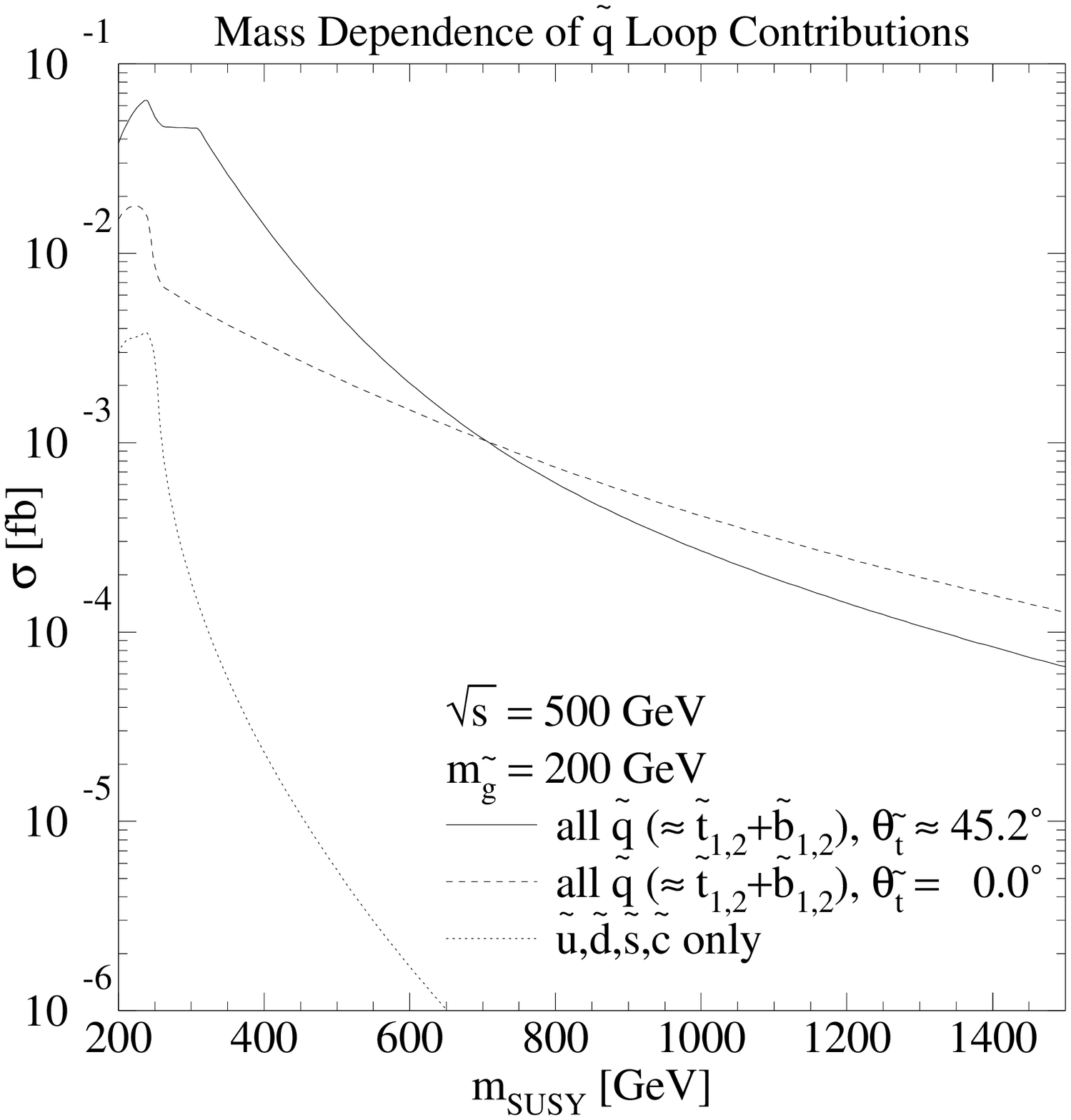,width=0.8\columnwidth}
 \caption{\label{fig:7}Squark mass dependence of the loop contributions from
 third generation squarks with maximal (full curve) and vanishing mixing
 (dashed curve). The contributions from the first two generations (dotted
 curve) are highly suppressed.}
\end{figure}

When $\msusy$ and the diagonal elements of the squark mixing matrix (see App.\
\ref{sec:a}) become much larger than the quark masses and the off-diagonal
elements of the matrix, the role of squark mixing is expected to be reduced.
This is confirmed numerically in Fig.\ \ref{fig:7}, where the dependence of
%
%
the gluino production cross section on $\msusy$ is shown for the cases of
maximal and vanishing top squark mixing. Squarks from the first two generations
contribute at most 10\% at low $\msusy$ and are otherwise strongly suppressed.

\begin{figure}
 \centering
 \epsfig{file=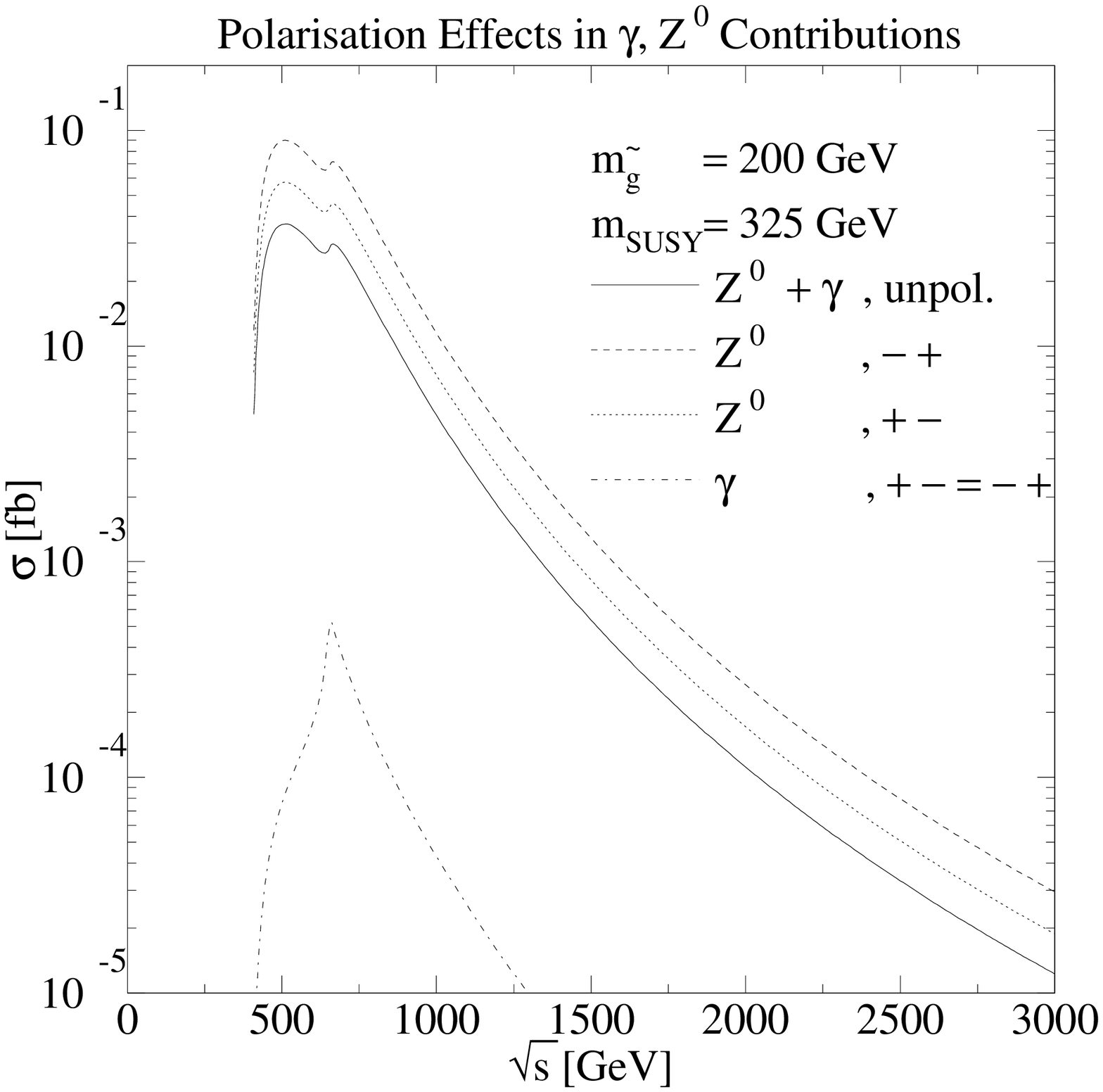,width=0.8\columnwidth}
 \caption{\label{fig:8}Center-of-mass energy dependence of the process
 $e^+e^-\to\go\go$ for unpolarized (full curve) and polarized (dashed and
 dotted curves) incoming electrons/positrons and maximal top squark mixing.
 The photon contribution (dot-dashed curve) is suppressed by more than two
 orders of magnitude.}
\end{figure}

At future linear $e^+e^-$ colliders it will be possible to obtain relatively
high degrees of polarization, {\it i.e.} about 80\% for electrons and 60\% for
positrons \cite{Aguilar-Saavedra:2001rg}. In Fig.\ \ref{fig:8} we therefore
%
%
investigate the effect of choosing different electron/positron polarizations
on the gluino pair production process, including contributions from all
(s)quarks. Since the $++$ and $--$ helicity amplitudes vanish for both photons
and $Z^0$-bosons, we only show the squares of the remaining $+-$ and $-+$
amplitudes, which coincide for photons, but not for $Z^0$-bosons. The
unpolarized cross section falls short of the polarized ones, so that a high
degree of polarization is clearly desirable.

\begin{figure}
 \centering
 \epsfig{file=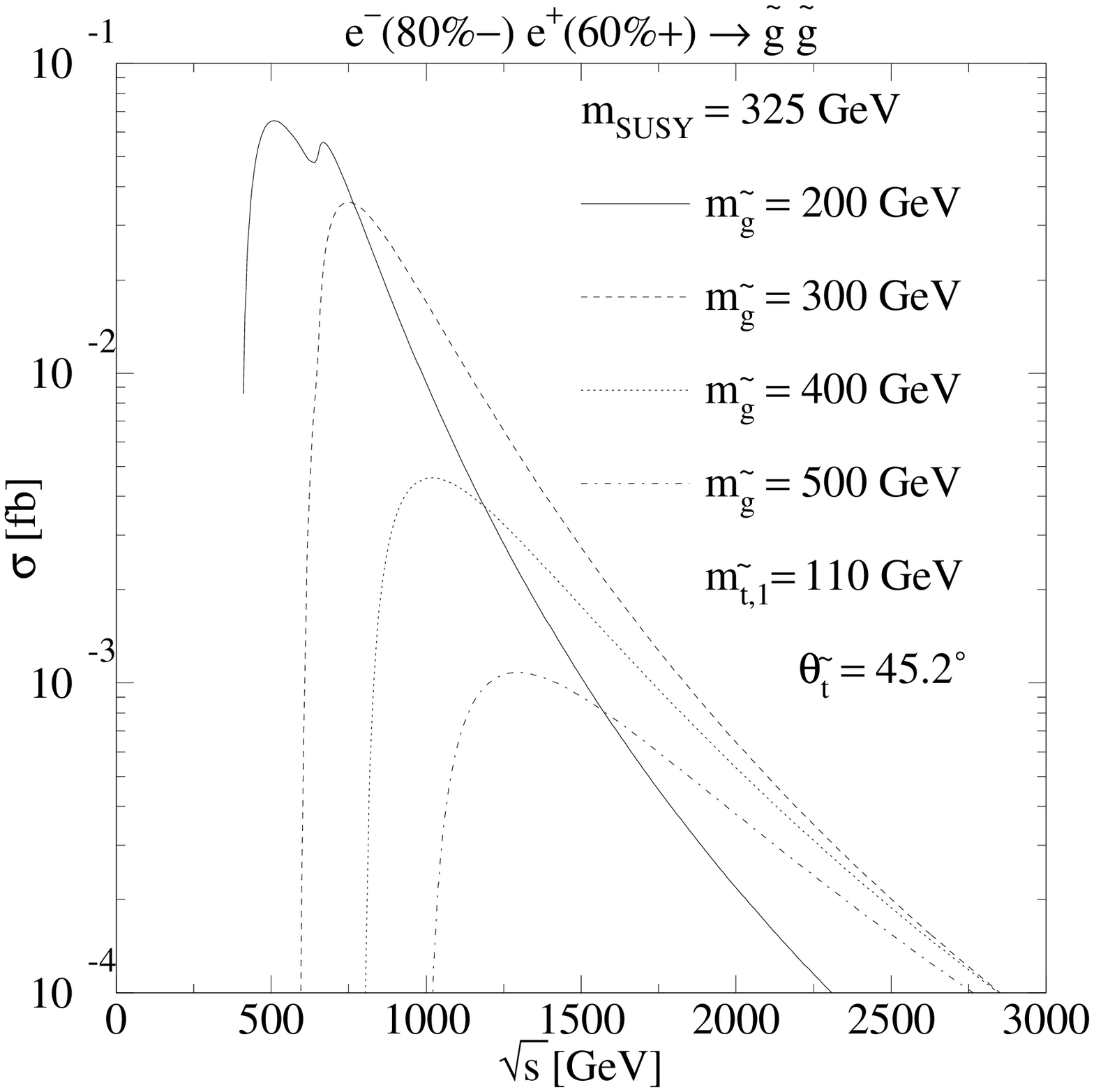,width=0.8\columnwidth}
 \caption{\label{fig:9}Center-of-mass energy dependence of the polarized
 $e^+e^-\to\go\go$ cross section for various gluino masses and maximal
 top squark mixing.}
\end{figure}

With the realistic degrees of polarization mentioned above, we show in
Fig.\ \ref{fig:9} a scan in the center-of-mass energy of a future $e^+e^-$
%
%
collider for various gluino masses and maximal top squark mixing (case II).
The cross section rises rather slowly due to the factor $\beta^3$ in Eq.\
(\ref{eq:polxsec}) for $P$-wave production of the gluino pairs. For $\mg=200$
GeV we observe an interesting second maximum, which arises from the
intermediate squark pair resonance at $\sqrt{s}=2\,\msusy=650$ GeV. At
threshold, the cross section depends strongly on the gluino mass and is largest
for $\mg= 200$ GeV, which we consider to be the lowest experimentally allowed
value. It drops fast with $\mg$, so that for $\mg>500$ GeV no events at
colliders with luminosities of 1000 fb$^{-1}$ per year can be expected,
irrespective of their energy. Smaller squark mixing (cf.\ Fig.\ \ref{fig:6})
or larger values of $\msusy$ (cf.\ Fig.\ \ref{fig:7}) will reduce the cross
section even further. Far above threshold, it drops off like $1/s$ and
becomes independent of the gluino mass.

\begin{figure}
 \centering
 \epsfig{file=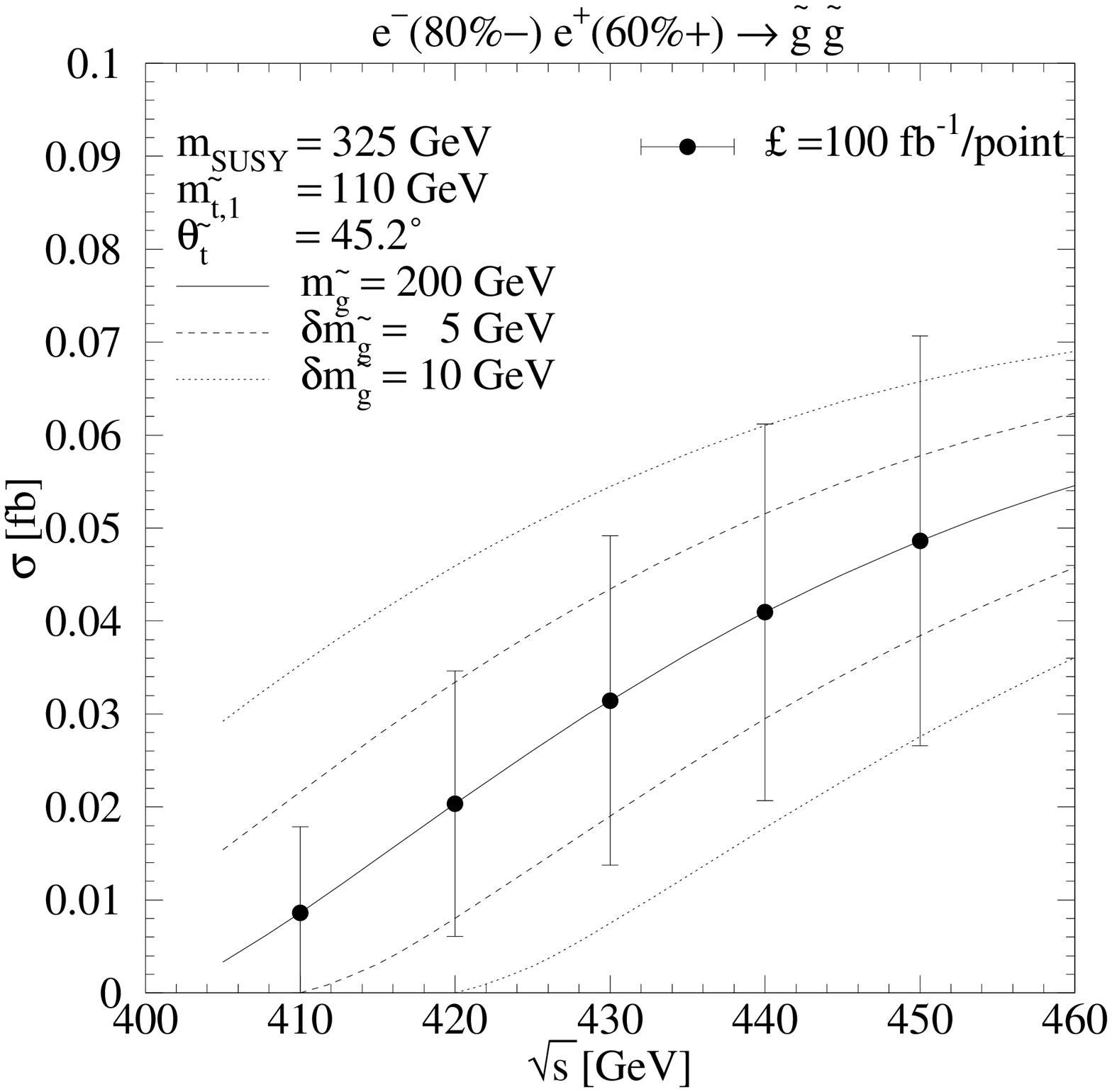,width=0.8\columnwidth}
 \caption{\label{fig:11}Sensitivity of the polarized $e^+e^-\to\go\go$ cross
 section to the gluino mass $\mg$ for maximal top squark mixing. The central
 values and statistical error bars of the data points have been calculated
 assuming $\mg=200$ GeV and a luminosity of 100 fb$^{-1}$ per center-of-mass
 energy point.}
\end{figure}

The slow rise of the cross section can be observed even better in Fig.\
\ref{fig:11}, where the sensitivity of a $\sqrt{s}=500$ GeV collider like
%
%
DESY TESLA to gluino masses around 200 GeV has been plotted. For the CERN LHC
experiments, a precision of $\pm30\,...\,60$ ($12\,...\,25$) GeV is expected
for gluino masses of 540 (1004) GeV \cite{:1999fr,Abdullin:1998pm}. If the
masses and mixing angle(s) of the top (and bottom) squarks are known, a
precision of $\pm5\,...\,10$ GeV can be achieved at DESY TESLA for $\mg=200$
GeV and maximal top squark mixing with an integrated luminosity of 100
fb$^{-1}$ per center-of-mass energy point.

\begin{figure}
 \centering
 \epsfig{file=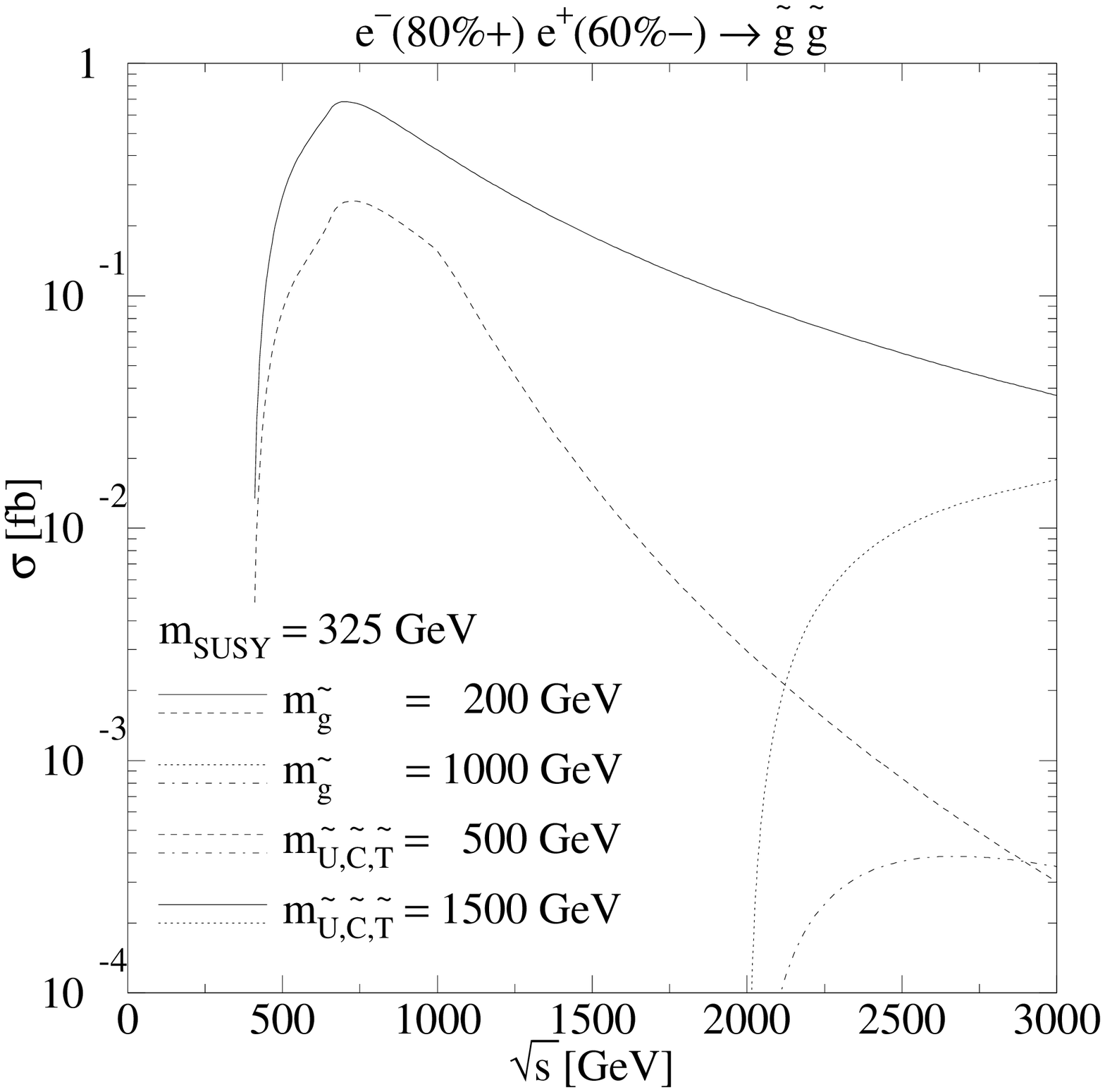,width=0.8\columnwidth}
 \caption{\label{fig:10}Center-of-mass energy dependence of the polarized
 $e^+e^-\to\go\go$ cross section for various gluino masses and mass splittings
 between left- and right-handed up-type squarks.}
\end{figure}

A center-of-mass energy scan for the scenario with no squark mixing,
but large left-/right-handed squark splitting (case I) is shown in Fig.\
\ref{fig:10} for
%
%
light (full and dashed curves) and heavy (dotted and dot-dashed curves) gluino
masses. Since the photon contributes now significantly to the cross section,
it proves to be advantageous to choose the lepton polarization such that the
$Z^0$-boson interferes constructively with the photon, even though it is by
itself slightly smaller than for the opposite choice. Since all three
generations add now to the cross section, it can become almost an order of
magnitude larger
than in the mixing scenario (case II), and even gluino masses of 1 TeV may
be observable at a multi-TeV collider like CERN CLIC. However, also here
the cross section drops sharply when the squark mass splitting is reduced
from 1500 TeV (full and dotted curves) to values close to $\msusy$ (dashed and
dot-dashed curves).

\section{Conclusions}
\label{sec:4}

The purpose of this Paper has been two-fold: First, we have resolved a
long-standing discrepancy in the literature about the relative sign of the
quark and squark loop contributions to the production of gluino pairs in
$e^+e^-$ annihilation. We confirm the result of two older papers
that the divergence cancels for each squark flavor separately and not between
weak isospin partners \cite{Kane:xp,Campbell:kf} and trace the sign problem in
one case to the Feynman rules employed in the corresponding calculation
\cite{Kileng:1994kw}. Our results rely on two completely independent analytical
calculations and one computer algebra calculation.

Second, we have investigated the prospects for precision measurements of
gluino properties, such as its mass or its Majorana fermion nature, at future
linear $e^+e^-$ colliders. We have taken into account realistic beam
polarization effects, photon and $Z^0$-boson exchange, and current mass
exclusion limits. Previously, only light gluinos at center-of-mass energies up
to the $Z^0$-boson mass had been investigated. Within the general framework of
the MSSM, we have concentrated on two scenarios of large left-/right-handed
up-type squark mass splitting and large top squark mixing, which produce
promisingly large cross sections for gluino masses up to 500 GeV or even
1 TeV. Gluino masses of 200 GeV can then be measured with a precision of about 
5 GeV in center-of-mass energy scans with luminosities of 100 fb$^{-1}$/point.
However, when both the left-/right-handed squark mass splitting and the
squark mixing remain small, gluino pair production in $e^+e^-$ annihilation
will be hard to observe, even with luminosities of 1000 fb$^{-1}$/year.


\begin{acknowledgments}

We thank B.\ Kileng, G.\ Kramer, and P.\ Osland for useful discussions and
A.\ Brandenburg and B.A.\ Kniehl for a careful reading of the manuscript.
This work has been supported by Deutsche Forschungsgemeinschaft through
Grant No.\ KL~1266/1-2 and through Graduiertenkolleg {\it Zuk\"unftige
Entwicklungen in der Teilchenphysik}.

\end{acknowledgments}

\appendix
\section{Squark Mixing}
\label{sec:a}

The (generally complex) soft SUSY-breaking terms $A_q$ of the trilinear
Higgs-squark-squark interaction and the (also generally complex) off-diagonal
Higgs mass parameter $\mu$ in the MSSM Lagrangian induce mixings of the left-
and right-handed squark eigenstates $\tilde{q}_{L,R}$ of the electroweak
interaction into mass eigenstates $\tilde{q}_{1,2}$.
The squark mass matrix \cite{Haber:1984rc,Gunion:1984yn}
\beq
 {\cal M}^2 =
 \lr\begin{array}{cc}
  m_{LL}^2+m_q^2  &
  m_q m_{LR}^\ast \\
  m_q m_{LR}      &
  m_{RR}^2+m_q^2
 \end{array}\rr
\eeq
with
\bea
 m_{LL}^2&=&(T_q^3-e_q\sin^2\theta_W)m_Z^2\cos2\beta+m_{\tilde{Q}}^2,\\
 m_{RR}^2&=&e_q\sin^2\theta_W m_Z^2\cos2\beta+\left\{\begin{array}{l}
 m_{\tilde{U}}^2\hspace*{4.8mm}{\rm for~up-type~squarks},\\
 m_{\tilde{D}}^2\hspace*{4.5mm}{\rm for~down-type~squarks},\end{array}\right.\\
 m_{LR}  &=&A_q-\mu^\ast\left\{\begin{array}{l}
 \cot\beta\hspace*{5mm}{\rm for~up-type~squarks}\\
 \tan\beta\hspace*{4.5mm}{\rm for~down-type~squarks}\end{array}\right.
\eea
is diagonalized by a unitary matrix $S$, $S {\cal M}^2 S^\dagger={\rm diag}\,
(m_1^2,m_2^2)$, and has the squared mass eigenvalues
\beq
 m_{1,2}^2=m_q^2+{1\over 2}\lr m_{LL}^2+m_{RR}^2\mp\sqrt{(m_{LL}^2-m_{RR}^2)^2
 +4 m_q^2 |m_{LR}|^2}\rr.
\eeq
For real values of $m_{LR}$, the squark mixing angle $\theta_{\tilde{q}}$,
$0\leq\theta_{\tilde{q}}\leq\pi/2$, in
\beq
 S = \lr \begin{array}{cc}~~\,\cos\theta_{\tilde{q}} &
                              \sin\theta_{\tilde{q}} \\
                             -\sin\theta_{\tilde{q}} &
                              \cos\theta_{\tilde{q}} \end{array} \rr
 \hspace*{3mm} {\rm with} \hspace*{3mm}
   \lr \begin{array}{c} \tilde{q}_1 \\ \tilde{q}_2 \end{array} \rr =
 S \lr \begin{array}{c} \tilde{q}_L \\ \tilde{q}_R \end{array} \rr
\eeq
can be obtained from
\beq
 \tan2\theta_{\tilde{q}}={2m_qm_{LR}\over m_{LL}^2-m_{RR}^2}.
\eeq
If $m_{LR}$ is complex, one may first choose a suitable phase rotation
$\tilde{q}_R'=e^{i\phi}\tilde{q}_R$ to make the mass matrix real and then
diagonalize it for $\tilde{q}_L$ and $\tilde{q}_R'$. $\tan\beta=v_u/v_d$
is the (real) ratio of the vacuum expectation values of the two Higgs fields,
which couple to the up- and down-type (s)quarks. The weak isospin quantum
numbers for left-handed up- and down-type (s)quarks with hypercharge $Y_q=1/3$
are $T_q^3=\{+1/2,-1/2\}$, whereas $Y_q=\{4/3,-2/3\}$ and $T_q^3=0$ for
right-handed (s)quarks, and their fractional electromagnetic charges are
$e_q=T_q^3+Y_q/2$.
The soft SUSY-breaking mass terms for left- and right-handed squarks
are $m_{\tilde{Q}}$ and $m_{\tilde{U}}$, $m_{\tilde{D}}$, respectively,
and $m_Z$ is the mass of the neutral electroweak gauge boson $Z^0$.

\section{Feynman Rules}
\label{sec:b}

Denoting squark mass eigenstates by $i,\,j,\,...$, Lorentz indices by
$\mu,\,\nu,\,...$, and color indices of the fundamental (adjoint)
representation of the color symmetry group SU(3) by $l,\,m,\,...$
$(a,\,b,\,...)$, we obtain the following propagators in Feynman-gauge:
\vspace*{-5mm}
%
\bea
 \epsfig{file=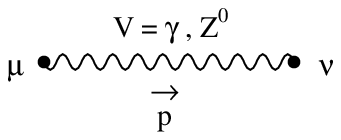,bbllx=55pt,bblly=390pt,bburx=165pt,bbury=430pt,%
  width=4cm} && {-i g^{\mu\nu} \over p^2-m_V^2+i\eta}
  \ , \  m_V^2 = \{0,m_Z^2\} \\
 \epsfig{file=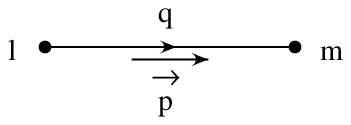,bbllx=55pt,bblly=390pt,bburx=165pt,bbury=430pt,%
  width=4cm} && { i (p\!\!\!/+m_q) \delta_{lm} \over p^2-m_q^2+i\eta} \\
 \epsfig{file=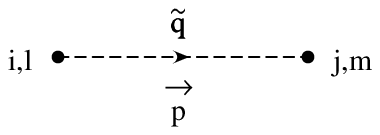,bbllx=55pt,bblly=390pt,bburx=165pt,bbury=430pt,%
  width=4cm} && { i \delta_{ij}\delta_{lm} \over p^2-m_{\tilde{q}}^2+i\eta}
\eea
%
Dirac fermions carry an arrow, which indicates the fermion number flow, whereas
Majorana fermions, such as gluinos, do not carry arrows. An additional
arrow is depicted next to all fermion lines in order to obtain a unique
orientation of the fermion flow, which is evaluated according to the rules
in \cite{Denner:1992vz}. The interaction vertices are given by
\cite{Haber:1984rc}
\vspace*{-15mm}
%
\bea
 \epsfig{file=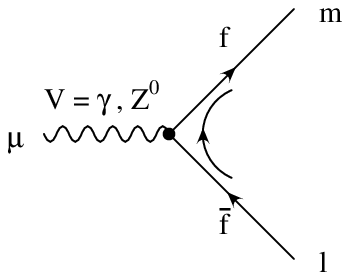,bbllx=55pt,bblly=390pt,bburx=165pt,bbury=480pt,%
  width=4cm} && -i e \gamma_\mu (v_f^V - a_f^V \gamma_5) \delta_{lm} \left\{
  \begin{array}{ll}
   v^\gamma_f = e_f &
   a^\gamma_f = 0 \\[8pt]
   v^Z_f      = {T_f^3-2 e_f \sin^2\theta_W\over
                 2 \sin\theta_W \cos\theta_W} &
   a^Z_f      = {T_f^3                     \over
                 2 \sin\theta_W \cos\theta_W}
  \end{array}
  \right. \label{eq:vert_yzqq} \\
 \epsfig{file=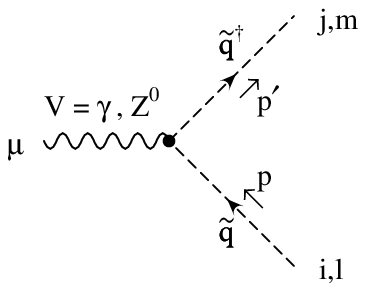,bbllx=55pt,bblly=390pt,bburx=165pt,bbury=480pt,%
  width=4cm} && -i e (p+p')_\mu \Gamma^{ij,V}_q \delta_{lm} \left\{
  \begin{array}{lcl}
   \Gamma^{ij,\gamma}_q \!\!\!&=&\!e_q \delta_{ij} \\[8pt]
   \Gamma^{ij,Z^0   }_q \!\!\!&=&\!{(T_q^3-e_q\sin^2\theta_W)S_{j1} S_{i1}^\ast
                                          -e_q\sin^2\theta_W S_{j2} S_{i2}^\ast
                                    \over\sin\theta_W\cos\theta_W}
  \end{array}
  \right. \\
 \epsfig{file=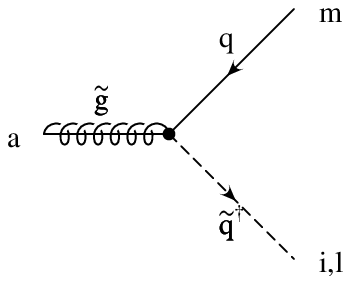,bbllx=55pt,bblly=390pt,bburx=165pt,bbury=480pt,%
  width=4cm} && i \Gamma_{i,1}^a = -i \sqrt{2} \hat{g}_s T^a_{lm}
  (S_{i1}      P_L-S_{i2}      P_R) \\
 \epsfig{file=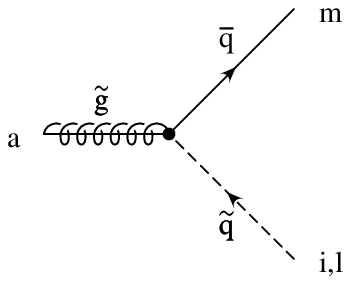,bbllx=55pt,bblly=390pt,bburx=165pt,bbury=480pt,%
  width=4cm} && i \Gamma_{i,2}^a = -i \sqrt{2} \hat{g}_s T^a_{ml}
  (S_{i1}^\ast P_R-S_{i2}^\ast P_L)
\eea
%

\vspace*{10mm}
\noindent
The Feynman rules in Eq.\ (\ref{eq:vert_yzqq}) apply to photon and $Z^0$-boson
interactions with quarks and charged leptons. The latter carry electromagnetic
charge $e_\ell=-1$ and weak isospin $T_\ell^3=-1/2$ (left-handed) and 0
(right-handed), but no color ($\delta_{lm} \to 1$).
The gauge couplings $g$ and $g'$ of the weak isospin and hypercharge
symmetries SU(2)$_{\rm L}$ and U(1)$_{\rm Y}$ have been expressed in terms
of the electromagnetic coupling $e=g\sin\theta_W$ and the sine and cosine of
the weak mixing angle $\sin\theta_W/\cos\theta_W=g'/g$.
The gluino-quark-squark vertices depend on the generators of the SU(3) color
symmetry group, $T^a_{lm}$, and on the Yukawa coupling $\hat{g}_s$, which is
identical to the strong gauge coupling $g_s$ in leading order, and on the
squark mixing matrix $S$, but not on the orientation of the fermion flow.



\end{document}